\documentclass[11pt]{article}
\usepackage{amsfonts}
\usepackage{amssymb}
\usepackage{graphicx}
\usepackage{amsmath}
\usepackage{amsthm}
\usepackage{color}
\usepackage{multirow}
\usepackage{url}
\usepackage{float}
\usepackage{booktabs} 
\usepackage{makecell}
\usepackage{listings}
\usepackage{subcaption}
\usepackage{algorithm}
\usepackage{enumerate}
\usepackage{tikz}

\usepackage{algpseudocode}
\lstnewenvironment{Rcode}{\lstset{language=R}}{}

\definecolor{battleshipgrey}{rgb}{0.52, 0.52, 0.51}
\definecolor{navyblue}{rgb}{0.0, 0.0, 0.5}
\definecolor{arsenic}{rgb}{0.23, 0.27, 0.29}
\definecolor{oldmauve}{rgb}{0.4, 0.19, 0.28}
\usepackage[colorlinks=TRUE]{hyperref}
\hypersetup{
	colorlinks=true,
	linkcolor=navyblue,
	filecolor=blue,      
	urlcolor=oldmauve,
	citecolor=navyblue,
	pdftitle={Overleaf Example},
	pdfpagemode=FullScreen,
}
\usepackage{natbib}
\setcitestyle{authoryear}

\begin{document}

	\title{ \bf Variable selection in functional regression models: a review}
	\author{Germ\'{a}n Aneiros{$^a$}\footnote{Corresponding author email address: \href{ganeiros@udc.es}{ganeiros@udc.es}} \hspace{2pt} Silvia Novo{$^b$} \hspace{2pt} Philippe Vieu{$^c$} \\		
		{\normalsize $^a$ Department of Mathematics, MODES, CITIC, ITMATI, Universidade da Coruña, A Coruña, Spain}\\
		{\normalsize $^b$ Department of Mathematics, MODES, CITIC, Universidade da Coruña, A Coruña, Spain}\\
		{\normalsize $^c$ Institut de Math\'{e}matiques, Universit\'e  Paul Sabatier, Toulouse, France}
	}

\date{}
\maketitle
\begin{abstract} 
	Despite of various similar features, Functional Data Analysis and High-Dimensional Data Analysis are two major fields in Statistics that grew up recently almost independently one from each other. The aim of this paper is to propose a survey on methodological advances for variable selection in functional regression, which is typically a question for which both functional and multivariate ideas are crossing. More than a simple survey, this paper aims to promote even more new links between both areas.
\end{abstract}


\section{Introduction}\label{sec:1}

Nowadays, Functional Data Analysis (FDA) is among the main fields in Statistics. The rich production is confirmed in various surveys (see, for instance, \citet{GoiaVieu2016}, \citet{Aneirosetal2019}). In the beginning, the presence of functional data in applications was rare. However, with the development of modern technology most applied sciences have to treat datasets containing one, or more, functional object. For the same reasons one has to treat High-(but finite) Dimensional Data, and High-Dimensional Statistics (HDS) grew up at the same time with FDA. The main common feature of both fields is that they take part of the recent enfatuation for Big Data Analysis. At the beginning, both fields developed in the statistical community in rather independent ways but the benefits that one could get by crossing ideas from both fields have been highlighted in the last decade as well in the HDS community (see \citet{ahmed2017}, \citet{Sangalli2018}, \citet{Vieu2018}) as in the FDA community (see \citet{aneiros2019}, \citet{bongiorno2014}). Following its 50 years long tradition of publishing top level innovative methodological advances on multidimensional data analysis, the Journal of Multivariate Analysis has played a leading role in the last decade for bridging gaps between FDA and HDS. This is, for instance, attested by
various special issues aiming to promote methodological advances by linking both fields (see \citet{GoiaVieu2016}, \citet{Aneirosetal2019}, \citet{Aneirosetal2022}). This paper aims to  celebrate this 50th birthday by proposing a review on variable selection methods within a functional framework which  is a topic where both fields are crossing in a natural way.

When dealing with a regression problem with one (or more) functional predictor, sometimes with some non-functional multivariate predictor, one has many concerns. First of all one has  to take into account the fact that we are
dealing with infinite-dimensionality of functional objects (see, for instance, \citet{Cuevas2014}) and to keep in mind the necessity of building models balancing flexibility, dimension reduction properties and interpretability (\citet{Vieu2018}). Secondly, one also has to worry about the quantity of information to be included into the model: this concerns the number of predictors as well as the number of discretizations that one has at hand for each functional predictor.  In a pragmatic way, as it is the case in HDS with non-functional high-dimensional predictors, one would like  to determine a smaller subset of variables that exhibits the strongest effects on the response (see \citet{Hastieetal2009}). %Being able to select the relevant information has became other of the important matters.
In the last decade, there has been a rather large production on sparse modelling and variable selection techniques in functional setting, and this article is aiming to review the state of art on this topic.

Our paper is organized as follows. Because most of the variable selection procedures in functional setting were extended from finite-dimensional regression, we start in Section \ref{Section1} with a selected review on the techniques employed in HDS with main attention on penalized methods. In  the exposition we will present the procedures by splitting them according to the nature of the model: linear, grouped and additive regression. Of course, Section \ref{Section1} is not supposed to be an exhaustive review of the very wide set of contributions existing in a multivariate setting, but only a presentation of those contributions which have been adapted for FDA. In Section \ref{Section2} we will go through the functional setting. The rich production and the variability in types of models, variables included and tools, led us to make distinction between four types of methodologies.  Subsections \ref{Section2.1}, \ref{Section2.2} and \ref{Section2.3} are dedicated to scalar response models which are most often studied in the literature: firstly we will study selection of scalar variables in models  which contain some functional predictor, secondly we will deal with the selection of scalar variables derived from the discretization of a functional object, and thirdly with the selection of functional objects. Finally, Subsection \ref{Section2.4} concerns the regression models with functional response. To conclude the paper, in Section \ref{Section3} we will present some  ideas about how variable selection could behave in functional regression in the next following years.

To make simpler the exposition of all the methodologies, we will assume without loss of generality that the involved variables (functional or not) are centred to have zero mean. In the same way, we will not show the assumptions (neither on the random errors nor the covariates) used in the different methodologies (note that such assumptions could change from one methodology to other one). 

\section{Variable selection in finite-dimensional regression  models}\label{Section1}

In the finite-dimensional setting, there is an extensive literature in variable selection tools (see, for instance, \citet{FanLv2010} or \citet{Desboulets2018} for recent reviews). In Section \ref{Section1}, we will present and briefly comment some of these techniques, paying main attention to those that have been extended to the functional framework. In that way, we could refer to them in the next section dedicated to functional models.

The production in variable selection procedures for finite-dimensional regression started with naive ideas such as stepwise regression (backward (\citet{Efroymson1960}), forward (\citet{Weisberg1980}) or both), forward-stagewise regression or best subset regression (\citet{FurnivalWilson1974}).
However, these methods are computationally intensive, unstable (see \citet{Breiman1996} or \citet{FanLi2001}) and it is hard to derive sampling properties. They are ``discrete procedures'' (variables are either selected or discarded), so they often exhibit high variance, and therefore, in some cases they do not reduce the prediction error of the full model.

For that, other techniques appeared, like shrinkage methods, also known as regularization, penalty-based or penalized methods. Shrinkage procedures are more continuous, and do not suffer as much from high
variability (see \citet{Hastieetal2009}). Most of these procedures attempt to select variables automatically and simultaneously (a notorious exception is bridge regression  for $L_q$ norms with $q>1$; see \citet{FrankFriedman1993} and \citet{FanLi2001}). These methods are based on adding a penalization term in the estimation task which, under suitable conditions, generates a sparse solution, in the sense that some estimated coefficients are zero. 
Penalized methods are highly developed, specially in the case of linear modelling. Specifically, the well-known linear model is given by the expression
$Y_i=\sum_{j=1}^p\beta_jX_{ij}+\varepsilon_i, \ i\in\{1,\dots,n\},$
where $Y_i$ is a scalar response, $\pmb{X}_i=(X_{i1},\dots,X_{ip})^{\top}$ is a vector of scalar covariates, $\pmb{\beta}=(\beta_1,\dots,\beta_p)^{\top}$ is a vector of unknown real coefficients and $\varepsilon_i$ is the random error.
The penalized estimator of the vector of unknown parameters, is the solution of the optimization problem
\begin{equation}
	\hat{\pmb{\beta}}=\arg\min_{\pmb{\beta}\in\mathbb{R}^p}\left( \ell(\pmb{\beta})+nP_{\lambda}(\pmb{\beta})\right),\label{beta_penalized_linear}
\end{equation}
where $\ell(\cdot)$ is a real-valued function which depends on the model and on its estimation procedure; if the estimation is made though penalized least squares, then 
$\ell({\pmb{\beta}})=(\pmb{Y}-\pmb{X}\pmb{\beta})^{\top}(\pmb{Y}-\pmb{X}\pmb{\beta})$, %P_{\lambda}(\pmb{\beta})\right).\label{beta linear}
where  $\pmb{Y}=(Y_1,\dots,Y_n)^{\top}$ and  $\pmb{X}=(\pmb{X}_1,\dots,\pmb{X}_n)^{\top}$. 
$P_{\lambda}(\cdot)$ is a penalty function which depends on a regularization parameter $\lambda>0$. The parameter $\lambda$ controls the amount of penalty and, in the case that (\ref{beta_penalized_linear}) gives rise to sparse solutions, also controls the sparseness of the resulting vector (as noted above, not all the estimators verifying (\ref{beta_penalized_linear}) give rise to sparse solutions).

The penalty function employed has a big influence to the properties of the derived estimator (see \citet{FanLi2001}). In the literature there are several proposals for this penalization term, but we are going to comment briefly the ones most used in the functional setting. Among penalty functions, the majority of them are based on norms.
Probably the most famous shrinkage method, based on norms, was proposed in \citet{Tibshirani1996}, where $L_1$ penalty was used:  \begin{equation}P_{\lambda}(\pmb{\beta})=\lambda \sum_{j=1}^p \left\lvert\beta_j\right\lvert. \label{L_1 penalty}\end{equation} He gave the name least absolute shrinkage and selection
operator (LASSO) method to the combination of this penalty with the least squares procedure. However, several objections emerged about this penalty.
On the one hand, LASSO estimators do not
satisfy oracle properties (see \citet{FanLi2001}). On the other hand, \citet{MeinshausenBuhlmann2006} showed that in LASSO the optimal $\lambda$ for prediction gives inconsistent variable selection results. This problem was also found by \citet{Lengetal2006}, who, in particular, showed that for any sample size $n$, when there are non relevant variables in the
model and the design matrix is orthogonal, the probability that LASSO correctly identify the true set of important variables is less than a constant (not depending on $n$) smaller than one. For that, other penalties were studied.
\citet{Zou2006} proposed adaptive LASSO (adaLASSO), where the  penalty term has the form \begin{equation}P_{\lambda}(\pmb{\beta})=\lambda \sum_{j=1}^p w_j\lvert\beta_j\lvert,\label{adaptive lasso penalty}\end{equation} and $w_j, \ j\in\{1,\dots,p\},$ are known weights. They showed that  if the weights
are data-dependent and cleverly chosen, then the adaptive LASSO estimators
can have the oracle properties. Another famous proposal is the elastic-net penalty (see \citet{ZouHastie2005}) which is a compromise between $L_1$ and $L_2$ penalties: 
$P_{\lambda}(\pmb{\beta})=\lambda_2\sum_{j=1}^p\lvert\beta_j\rvert+\lambda_1\sum_{j=1}^p\beta_j^2$. In a general way,
\citet{Huangetal2008} studied bridge penalties $P_{\lambda}(\pmb{\beta})=\lambda \sum_{j=1}^p\lvert\beta_j\lvert^{q}$ (related with the $L_q$ norm) and showed that they verify the oracle property for $0<q<1$. In that paper they consider the more general context in which
the number of covariates, say $p_n$, may increase to infinity with $n$ ($p_n\rightarrow\infty$ as $n\rightarrow\infty$).
In addition, a robust approach was studied in \citet{Wang2etal2007}, where instead of least squares estimation, they used least absolute deviation (LAD) with $\ell({\pmb{\beta}})=\sum_{i=1}^n\lvert Y_i-\sum_{j=1}^p\beta_jX_{ij}\rvert$ combined with $L_1$ penalty (LAD-LASSO).

Probably the main competitor of  penalties based on norms is the proposal in \citet{FanLi2001}: the smoothly clipped absolute deviation penalty (SCAD) defined,  for $a>2$, as  
\begin{equation}
	P_{\lambda}(\pmb{\beta})=\sum_{j=1}^p\mathcal{P}_\lambda(\beta_j), \ \ \ \ \mathcal{P}_{\lambda}(\beta_j)=\left\{\begin{aligned}
		&\lambda\left|\beta_j\right|,&  \quad |\beta_j|<\lambda,\\
		&\frac{(a^2-1)\lambda^2-(|\beta_j|-a\lambda)^2}{2(a-1)},&\quad \lambda\leq |\beta_j|<a\lambda, \\
		&\frac{(a+1)\lambda^2}{2},&\quad |\beta_j|\geq a\lambda
	\end{aligned}
	\right.
	\label{SCAD}
\end{equation} 
(\citet{FanLi2001} suggested to take $a=3.7$).  
SCAD penalty improves properties of $L_1$ penalty, satisfying the oracle property. For that, it was very often used in works related with generalized linear models (GLM), in which was assumed that $Y_i$ is a real variable verifying $\mathbb{E}(Y_i|\pmb{X}_i)=g^{-1}(\eta_i)$ with  $\eta_i=\pmb{X}_i^{\top}\pmb{\beta} \ (i\in\{1,\dots,n\})$ and where % $\pmb{X}_i=(X_1,\dots,X_p)^{\top}$ is a vector of real covariates, $\pmb{\beta}=(\beta_1,\dots,\beta_p)$ is a vector of unknown coefficients and 
$g(\cdot)$ is a known injective continuous link function. 
\citet{FanLi2001} studied GLM and proposed obtaining a penalized log-likelihood estimator using SCAD. That is, the estimator derived from (\ref{beta_penalized_linear}) when $\ell(\pmb{\beta})$ denotes  the conditional log-likelihood of $Y_i$ and $P_{\lambda}(\pmb{\beta})$ is the SCAD (\ref{SCAD}). They studied properties of this estimator for fixed number of covariates $p$, while \citet{FanPeng2004} studied them when the number of covariates $p=p_n$ diverges ($p_n\rightarrow\infty$ as $n\rightarrow\infty$).

The extension of shrinkage methods to the context of grouped models (or multifactor analysis-of-variance (ANOVA) models; see \citet{YuanLin2006}) follows ideas that will be used later in functional variable selection. For that, we are going to include them in this brief revision. In these models each explanatory factor is represented by a group of derived input variables. Specifically, the grouped linear model is given by the relationship
\begin{equation}
	Y_i=\sum_{m=1}^M\pmb{X}_{im}^{\top}\pmb{\beta}_m+\varepsilon_i, \ i\in\{1,\dots,n\},\label{grouped linear model}
\end{equation}
where  regressors are divided into $M$ groups, so $\pmb{X}_{im}=(X_{im1},\dots,X_{imv_m})^\top$ (the case  $v_1=\dots=v_M=1$ gives standard linear regression) and  ${\pmb{\beta}}_m=(\beta_{m1},\dots,\beta_{mv_m})^{\top}, \ m\in\{1,\dots,M\}$.  Therefore, in this case, the interest is not in selecting variables individually, but in choosing important factors and each one  is in correspondence with a group of covariates. 
Therefore, the following optimization problem should be solved:
\begin{equation}
	\hat{\pmb{\beta}}^* =\arg\min_{\pmb{\beta }^{\ast }\in \mathbb{R}^{v_{1}}\times
		\cdots \times \mathbb{R}^{v_{M}}}\left(\ell^*(\pmb{\beta}^*)+nP_{\lambda}^*(\pmb{\beta}^*)\right),\label{beta group}
\end{equation}
where $\pmb{\beta}^*=(\pmb{\beta}_1,\dots,\pmb{\beta}_M)$ and $P_{\lambda}^*(\cdot): \mathbb{R}^{v_{1}}\times
\cdots \times \mathbb{R}^{v_{M}}\rightarrow\mathbb{R}$ denotes the penalty function.
Note that to obtain a penalized least squares estimator $\ell^*(\pmb{\beta}^*)=\sum_{i=1}^n \left(Y_i-\sum_{m=1}^M\pmb{X}_{mi}^{\top}\pmb{\beta}_m\right)^2$ should be considered. The question now is how to choose the penalty function for selecting groups of covariates.
\citet{YuanLin2006} proposed the group LASSO penalty defined as
\begin{equation}
	P_{\lambda}^*(\pmb{\beta}^*)=\lambda\sum_{m=1}^M\sqrt{\pmb{\beta}_m^{\top}K_m\pmb{\beta}_m}, \label{Group LASSO penalty}
\end{equation}
where  $K_m$ is a positive definite matrix, $m\in\{1\dots,M\}$. Penalty (\ref{Group LASSO penalty}) is intermediate between the $L_1$ and the $L_2$ penalties. A derived problem is the selection of the  matrices $K_m$; \citet{YuanLin2006} used $K_m=v_mI_{v_m}$ with $m\in\{1\dots,M\}$ where $I_{v_m}$ is the identity matrix of size $v_m$. %\citet{Meieretal2008} adapt group LASSO procedure for generalized grouped linear models.
The adaptation of the LASSO gave the way to other extensions, like the group SCAD penalty
\begin{equation}
	P_{\lambda}^*(\pmb{\beta}^*)=\lambda\sum_{m=1}^M\mathcal{P}_{\lambda}\left(\sum_{r=1}^{v_m}\beta_{mr}^{2}\right), \label{Group SCAD penalty}
\end{equation}
where $\mathcal{P}_{\lambda}(\cdot)$ was defined in (\ref{SCAD}).
This penalty was proposed in \citet{Wangetal2007} in the context of the varying coefficients models with functional response, that we will discuss later. These authors also proved oracle properties for this penalty.
Another general proposal suggests to use composite absolute penalties (CAP) studied in \citet{Zhaoetal2009}. CAP depend on a vector of norm parameters, $(\gamma_0, \gamma_1,\ldots,\gamma_M)$; these penalties are given by the expression
\begin{equation}
	P_{\lambda}^*(\pmb{\beta}^*)=\lambda\sum_{m=1}^M\lvert \lvert \lvert \pmb{\beta}_m\rvert\rvert_{\gamma_m}\rvert^{\gamma_0},\label{CAP}
\end{equation}
where $||\cdot||_\gamma$ denotes the $L_\gamma$ norm.
The parameter $\gamma_0$ determines how groups
relate to each other while $\gamma_m$ dictates the relationship of the coefficients within
group $m$. 
Therefore, this family of penalties allows grouped selection and the hierarchical variable selection is
reached by defining groups with particular overlapping patterns.

So far we have studied penalized methods for linear models. However, these procedures can be  employed even in nonlinear regression. An interesting case for the relations with functional regression is additive model given by the expression
$Y_i=\sum_{m=1}^Mf_m(X_{mi})+\varepsilon_i, \ i\in\{1,\dots,n\},$
where $f_m(\cdot)$ with $m\in\{1,\dots,M\}$ are smooth univariate real-valued functions which should be estimated. In this case the optimization problem is carried out in a space of  functions, say $\mathcal{F}$, since the target functions are the solution of
\begin{equation}
	\hat{\pmb{f}}^* =\arg\min_{\pmb{f}^*\in\mathcal{F}} \left(\ell^{**}(\pmb{f}^*) +nP_{\lambda}^{**}(\pmb{f}^*)\right), \label{beta additive models}
\end{equation}
where $\pmb{f}^*=(f_1,\dots,f_M)^{\top}$ and $P_{\lambda}^{**}(\cdot):\mathcal{F}\rightarrow\mathbb{R}$ is a penalization term. 
For these models, \citet{Meieretal2009} proposed $\ell^*(\pmb{f}^*)=\sum_{i=1}^n\left(Y_i-\sum_{m=1}^Mf_m(X_{mi})\right)^2$ and, as penalty function,
the sparsity-smoothness penalty that simultaneously controls smoothing of functions $f_m(\cdot)$  and sparseness, 
\begin{equation}
	P_{\lambda}^{**}(\pmb{f}^*)=\sum_{m=1}^M \mathcal{P}_{\lambda_1,\lambda_2}(f_m),  \ \ \ \mathcal{P}_{\lambda_1,\lambda_2}(f_m)=\lambda_1\sqrt{\frac{1}{n}\sum_{i=1}^n f_m(X_{im})^2+\lambda_2\int (f_m''(x)dx)^2}. \label{sparsity-smoothness penalty}
\end{equation}
Two tuning parameters $\lambda_1$ and $\lambda_2$ control the amount of penalization: $\lambda_1$ is a sparseness/tuning parameter and $\lambda_2$ is a smoothing/tuning parameter, since the second term in $\mathcal{P}_{\lambda_1,\lambda_2}(\cdot)$ controls the smoothness of functions $f_m(\cdot)$ with $m\in\{1,\dots,M\}$. To solve the optimization problem (\ref{beta additive models}) in practice, \citet{Meieretal2009} use cubic B-spline basis expansion of functions $f_m(\cdot)$, that is, $f_m(x)=\sum_{r=1}^V\beta_{mr}b_{mr}(x)$, where $b_{mr}(\cdot)$ are B-spline basis functions and $\pmb{\beta}_m=(\beta_{m1},\dots,\beta_{mV})^{\top}$ is the parameter vector of $f_m(\cdot)$.  
In this way, \citet{Meieretal2009} reduced the optimization problem (\ref{beta additive models}) to (\ref{beta group}), with $v_m=V$, since number of functions in the B-spline basis is independent from $m$, and penalty (\ref{sparsity-smoothness penalty}) adopts the form of the group LASSO penalty (\ref{Group LASSO penalty}).
\citet{Huangetal2010} also studied an adaptive group LASSO procedure for additive modelling.

Although we have focused the exposition on shrinkage methods, other different procedures have been proposed in the literature to select relevant variables. In the context of linear modelling  \citet{Efronetal2004} proposed a Least Angle Regression (LARS) algorithm, a refined version of the
forward stagewise procedure that uses a simple mathematical formula to accelerate
the computations. This method is computationally efficient  and it has  LASSO (LARS-LASSO) and forward stagewise methods as variants. 
A different idea is the Dantzig selector proposed in \citet{CandesTao2007}, based on linear programming, which is able to deal with the case $p \gg n$ (that is, $p$ is much larger than $n$). Another important contribution was the sure independence screening procedure proposed in \citet{FanLv2008}, based on correlations. The enumeration of methods could go on; see, for instance, \citet{Lietal2012} for a distance correlation method, \citet{Keetal2014} for sparse models where signals 
are both rare and weak, \citet{MielniczukTeisseyre2014} for a random subspace method and \citet{Oharaetal2009} for a review of Bayesian approaches.

\section{Variable selection in functional regression models}\label{Section2}
We have presented variable selection methods in the finite-dimensional context. Here we are going to study their extension to the infinite-dimensional setting and Section \ref{Section2} is the main part of our paper. Because variable selection may occur from various points of view,  we will divide the exposition into four subsections. The first three subsections are dealing with the scalar response: in  Section \ref{Section2.1} we are going to revise works dealing with scalar variable selection when models also contain functional predictors; in Section \ref{Section2.2} we will study variable selection of scalar covariates originated from the discretization of a curve; in Section \ref{Section2.3} we are going to deal with functional covariate selection. Finally, in Section \ref{Section2.4} we are going to treat models with functional response.

\subsection{Selection of scalar covariates}\label{Section2.1}
As commented in the introduction, the combination of scalar and functional predictors in applications becomes a frequent question in many applied sciences problems. 
One has situations where, in addition to a very large number of covariates, $p_n$, there is also some functional predictor involved.
\citet{AneirosFerratyVieu2015} dealt with this reality in the case of a scalar response, proposing a sparse partial linear model with functional covariate, which allows $p_n\rightarrow\infty$ as $n\rightarrow\infty$. The model that they studied is given by the expression
\begin{equation}
	Y_i=\pmb{X}_i^{\top}\pmb{\beta}+m(\zeta_i)+\varepsilon_i, \label{SSFPLM} \ i\in\{1,\dots,n\},
\end{equation}
where $Y_i$ is the scalar response, $\pmb{X}_i=(X_{i1}\dots, X_{ip_n})^{\top}$ are real random covariates, $\zeta_i=\zeta_i(t)$ is the functional random covariate valued in a semi-metric space and
$\pmb{\beta}=(\beta_{1},\dots,\beta_{p_n})^{\top}\in\mathbb{R}^{p_n}$ is the vector of unknown parameters, $m(\cdot)$ is the nonlinear unknown link operator and $\varepsilon_i$ is the random error. The strategy that they propose is to carry out variable selection in the linear component by transforming model (\ref{SSFPLM}) into a linear one. For that, the effect of the functional covariate should be extracted from the response and the other scalar predictors. That is, one should consider the model
\begin{equation}
	Y_i-\mathbb{E}(Y_i|\zeta_i)=\sum_{j=1}^{p_n}\beta_{j}(X_{ij}-\mathbb{E}(X_{ij}|\zeta_i))+\varepsilon_i, \ i\in\{1,\dots,n\}.\label{LMtransformation}
\end{equation}
For estimating the conditional expectations $\mathbb{E}(\cdot|\zeta_i)$ in the expression (\ref{LMtransformation}), functional nonparametric regression can be employed (see \citet{FerratyVieu2006}). Once the model is transformed (in an approximate way) into a linear one, penalized estimation (\ref{beta_penalized_linear}) can be applied. In \cite{FerratyVieu2006} the SCAD penalty (\ref{SCAD}) was used.

In the same context \citet{NovoAneirosVieu2020}  assumed semiparametric effect for the functional predictor. That is, $m(\zeta_i)=g(\left<\theta,\zeta_i\right>)$, where $\zeta_i$ belongs to a separable Hilbert space with inner product denoted by $\left<\cdot,\cdot\right>$,  $\theta=\theta(t)$ is an unknown functional parameter and $g(\cdot)$ is a real-valued smooth function to estimate. In this case, for the transformation into a linear model, conditional expectations in expression in (\ref{LMtransformation}) were estimated using functional single-index regression (see \citet{AitSaidietal2008}). Penalized least squares estimation with SCAD penalty  (\ref{SCAD}) were applied to the resulting model.

\subsection{Selection of scalar covariates with functional origin}\label{Section2.2}
Infinite-dimensionality of functional objects has constantly been of concern in the FDA literature. When FDA was still not developed, predictive modelling in applied areas consisted in considering the discretized functional object  $\mathcal{X}(t)$, that is, scalar variables with functional origin, $\mathcal{X}(t_1),\dots,\mathcal{X}(t_p)$. Then, existing techniques were applied (like principal components regression or partial least squares) to reduce the dimension (see \citet{FrankFriedman1993} for a review).
Since FDA emerged, other techniques were proposed in order to reduce dimensionality of the functional predictor, but taking into account its continuous nature. % to a low number of dimensions.
The concept of ``sparseness'' in functional regression was usually not assumed with respect to the coefficients of the model. The common practise was to rewrite the model using a ``sparse'' expansion of $\mathcal{X}(t)$ (see, for instance, \citet{RamsaySilverman2005}).  At this stage it is worth to stress that the word ``sparse'' is used in FDA for different purposes (see \citet{AneirosVieu2016} for a discussion): here we are meaning sparsity in the model and not for the curve data itself.

However, in some recent publications the interpretability of the results led authors back to consider discretized functional objects. In addition, they realized that discretized values of the curves  $\mathcal{X}(t_1),\dots,\mathcal{X}(t_p)$ may contain information
which is not reachable through the continuous curve $\mathcal{X}(t)$, and conversely. Then different modelling options emerge (such as \citet{McKeagueSen2010}), many of them combined with the sparse concept in finite-dimensional regression as we will see in this section. In this case, new proposed procedures for variable selection are designed to deal with the very strong dependency between resulting scalar variables (taking into account the continuous origin) and with the very-high-dimension of the resulting vector. In addition, these sparse ideas are combined with either parametric, nonparametric or semiparametric regression modelling. 

In \citet{FerratyHallVieu2010}, authors follow a nonparametric approach. Specifically, suppose that $Y_i$ is a scalar response variable and $\mathcal{X}_i(t)$  is a functional random predictor  with   $t\in\mathcal{I}$ and  $\mathcal{I}$ is a compact subset of the real line. 
The functional nonparametric model (FNM) is given by the expression:
\begin{equation}
	Y_i=m(\mathcal{X}_i)+\varepsilon_i,\label{FNM} \ i\in\{1,\dots,n\},
\end{equation}
where $m(\cdot)$ is a smooth functional and $\varepsilon_i$ denotes the random error (for details on this model, see \citet{FerratyVieu2006}). 
In \citet{FerratyHallVieu2010}, they studied how to select the most predictive design points of the curve $\mathcal{X}_i(t)$, say $t_1,\dots,t_s$, using a procedure based on local linear regression (properties and references about local linear regression can be found in \cite{FanGijbels1996}). For that, they consider the discretized $\mathcal{X}_i(t)$ and transform the functional model (\ref{FNM}) into the underlying multivariate
nonparametric model:
\begin{equation}
	Y_i=g(\mathcal{X}_i(t_1),\dots,\mathcal{X}_i(t_p))+\varepsilon_i,   \ i\in\{1,\dots,n\},\label{MNM}
\end{equation} where $g(\cdot): \mathbb{R}^p\rightarrow\mathbb{R}$.
Then, they transform the estimation of the most predictive design points into a
multivariate function estimation problem. For that, they propose a two stages algorithm based on the cross-validation (CV) function:
\begin{equation}
	cv(\pmb{t}^*,h)=\frac{1}{n}\sum_{i=1}^n (Y_i-\widehat{g}_{h,-i}(\mathcal{X}_i(\pmb{t}^*)))v(\mathcal{X}_i(\pmb{t}^*)),  \ i\in\{1,\dots,n\},\label{CVfunction}
\end{equation}
where $\widehat{g}_{h,-i}(\cdot)$ is the leave-one-out local linear estimator of $g(\cdot)$, $\pmb{t}^*$ is a vector of design points and $\mathcal{X}_i(\pmb{t}^*)$ denotes the discretized values of the functional object at these design points, $h$ is a vector of smoothing parameters and $v(\cdot)$ is a nonnegative, integrable function of $s$ variables, which allows cases of
marked heterocedasticity (under homocedasticy, $v(\mathcal{X}_i(\pmb{t}^*))$ is set to $1$).
\begin{itemize}
	\item In the first stage, called forward addition, the algorithm adds the most predictive design points (in correspondence with
	a criterion based on (16)) step by step, while the addition of such points diminishes the value of the PCV function (a penalized version of the CV function, which penalizes the number of selected points). 
	\item 
	In the second stage, named backward deletion, the algorithm deletes the least predictive points (in correspondence with
	a criterion based on (\ref{CVfunction})) step by step, while the elimination of such points  diminishes the value of the PCV. Note that the second step allows to enlarge the number of
	possible combinations and therefore to find lower values of the cross-validation criterion.

\end{itemize}
As exposed in \citet{FerratyHallVieu2010}, the algorithm could be combined with variable selection penalty methods (like LASSO or LARS) to preselect design points and accelerate calculations. In addition, a boosting step could be added to improve the predictive performance. In fact, authors conclude that the incorporation of both discrete and continuous aspects of functional predictors could benefit the prediction power.

In \citet{KneipSarda2011}, authors follow a parametric approach. They also work with a scalar response variable $Y_i$ and a discretized functional covariate $\mathcal{X}_i(t)$. %(observed at an equidistant grid $t_j=1/p$)
In this case they consider the linear model
\begin{equation}Y_i=\sum_{j=1}^p\beta_j\mathcal{X}_{i}(t_j)+\varepsilon_i, \ i\in\{1,\dots,n\},\label{SLM1}\end{equation}
where $\pmb{X}_i=(\mathcal{X}_i(t_1),\dots,\mathcal{X}_i(t_p))^{\top}$ is  the discretized functional predictor, $\pmb{\beta}=(\beta_1,\dots,\beta_p)^{\top}$ is the vector of unknown parameters and $\varepsilon_i$ is the random error.
They studied variable selection in a linear factor model by assuming that the predictor $\pmb{X}_i$ can be decomposed into a sum of two uncorrelated random components in $\mathbb{R}^p$, 
\begin{equation*}
	\pmb{X}_i=\pmb{W}_i+\pmb{Z}_i, \ i\in\{1,\dots,n\},
\end{equation*}
where $\pmb{W}_i$ is intended to describe high correlations of the $X_{ij}=\mathcal{X}_i(t_j)$ while the components $Z_{ij}$ of $\pmb{Z}_i$, $j \in \{1,\ldots,p\}$, are uncorrelated. \citet{KneipSarda2011} assume that the components $W_{ij}=W_i(t_j)$ of $\pmb{W}_i$ as well as $Z_{ij}$ represent nonnegligible parts of the variance of $X_{ij}$ (common variability and specific variability, respectively). Taking the decomposition above into account, model (\ref{SLM1}) can be expressed as
\begin{equation}
	Y_i=\sum_{j=1}^p\beta_j^{**}W_{i}(t_j)+\sum_{j=1}^p\beta_j \mathcal{X}_{i}(t_j)+\varepsilon_i, \ i\in\{1,\dots,n\},\label{FLM}
\end{equation} where $\beta_j^{**}=\beta_j^{*}-\beta_j$ and
$
Y_i=\sum_{j=1}^p\beta_j^{*}W_{i}(t_j)+\sum_{j=1}^p\beta_jZ_{ij}+\varepsilon_i, \ i\in\{1,\dots,n\}.$ 

In order to estimate and select relevant variables in model (\ref{FLM}), \citet{KneipSarda2011} use different techniques to delete dependency between variables.
\begin{enumerate}
	\item On the one hand, variables $W_{i}(t_j), \ j\in\{1,\dots,p\},$ are heavily correlated. However, the term $\sum_{j=1}^p\beta_j^{**}W_{i}(t_j)$ could represent an important, common effect of all covariates. For avoiding the effect of the dependence, \citet{KneipSarda2011} propose to employ principal components to rewrite $\pmb{W}_{i}=\sum_{j=1}^p\pmb{\psi}_j^{\top}\pmb{W}_i\pmb{\psi}_j$. Then, they assume that the effect of $\pmb{W}_i$ can be described with a suitable small number of components, say $d$ ($d\leq p$). In that way, model (\ref{FLM}) takes the form
	\begin{equation}
		Y_i=\sum_{r=1}^d\alpha_r(\pmb{\psi}_r^{\top}\pmb{W}_i)+\sum_{j=1}^p\beta_j \mathcal{X}_{i}(t_j)+\varepsilon_i, \ i\in\{1,\dots,n\},\label{FLM2}
	\end{equation}
	where $\alpha_r=\sum_{j=1}^p\beta_j^{**}\psi_{jr}$.
	\item Before estimating and carrying out variable selection in model (\ref{FLM2}), the dependence between $\pmb{\psi}_r^{\top}\pmb{W}_i$ and $\mathcal{X}_{i}(t_j)$ should be deleted. For that, \citet{KneipSarda2011} used a projected model. They consider as predictor, instead of $\pmb{X}_i$, the projection of $\pmb{X}_i$ onto the orthogonal space of the  space spanned by the eigenvectors corresponding to the $k$ largest eigenvalues of the covariance matrix of $\pmb{X}_i$. %That is, using the projection matrix 
	Then, a variable selection procedure (such as  LASSO or the Dantzig selector) is applied to the resulting model to select relevant variables simultaneously in both components.
\end{enumerate}

\citet{Kneipetal2016} study a slightly different approach. They consider a generalization of the classical functional linear regression model assuming that there exists an unknown number of ``points of impact'', that
is, discrete observation times, where the corresponding functional values possess
some significant influences on the response variable.  Specifically, they consider the model
\begin{equation}
	Y_i=\int_{\mathcal{I}}\alpha(t)\mathcal{X}_i(t)dt+\sum_{k=1}^s\beta_k\mathcal{X}_i(t_k)+\varepsilon_i, \ i\in\{1,\dots,n\},
\end{equation}
where  $\mathcal{X}_i(t)$ is a curve with domain in the interval $\mathcal{I}$  and $\alpha(t)$ is a function parameter, while  $t_1,\dots,t_s\in\mathcal{I}$ are the points of impact where the curve has influence on the response.
For estimating the model, they should identify which discretized times of $\mathcal{X}_i(t)$ enter into the second component of the model, a topic related with variable selection of scalar covariates with functional origin. For estimating the number and location of impact points, they extract local variations from the functional covariate (to diminish correlations between $\mathcal{X}_i(t_k), \ k \in \{1,\ldots,s\}$), that is, define $Z(\mathcal{X},t)=\mathcal{X}(t)-(1/2)(\mathcal{X}(t-\delta)+\mathcal{X}(t+\delta))$ for $\delta>0$ with $[t-\delta,t+\delta]\in\mathcal{I}$ and choose as impact points those time points where there is a special high correlation between the response and $Z(\mathcal{X}_i,t)$.

A different idea for selection of impact points in (\ref{SLM1}) was proposed in \citet{Berrenderoetal2019}.
They assume that in the associated functional linear model, the function parameter, say $\alpha(t)$, belongs to a Reproducing Kernel Hilbert Space (RKHS) 
instead of the more usual $L_2$ space. Using the properties derived from such an assumption, they define an optimality criterion for selecting impact points which only depends on the covariance function of $\mathcal{X}(t)$ at each pair of time points and  on the covariance between $\mathcal{X}_i(t_j)$ and  $Y_i, \ j\in\{1,\dots p\}, \ i\in\{1,\dots n\}$. Based on the optimality criterion, they introduce a recursive expression that is used to carry out the selection.

\citet{AneirosVieu2014} follow a different approach for dealing with both dependency in the discretized curve and sparse linear modelling. In fact, their idea is to build a specific method for the case in which  scalar covariates have continuous origin.
They work with the linear model (\ref{SLM1})

where $Y_i$ is a scalar response and assume that $\mathcal{X}_i(t)$ is a random curve observed at the grid $a\leq t_1\leq \dots \leq t_{p_n}\leq b$, $\beta_j$ with $j\in\{1,\dots,p_n\}$ are the unknown coefficients and $\varepsilon_i$ the random error. Note that in this case $p=p_n$, that is, it is allowed that the discretization size tends to infinity with the sample size ($p_n\rightarrow\infty$ as $n\rightarrow\infty$). Therefore, for selecting relevant variables and estimating model (\ref{SLM1}) they propose the so-called partitioning variable selection (PVS) procedure. This two-stage algorithm relies on the idea that the values $\mathcal{X}(t_j)$ and $\mathcal{X}(t_k)$ with $t_j$ and $t_k$ very close will contain very similar information of the response. 
\begin{itemize}
	\item In the first stage, a reduced linear model is considered, with only very few covariates, say $w_n$, cover the entire discretization interval for $\mathcal{X}_i(t)$. That is, assuming without loss of generality that $p_n=q_nw_n$, the $w_n$ variables taken into account are $\mathcal{X}_i(t_{(2k-1)q_n/2}), \ k\in\{1,\dots,w_n\}$. The rest of the $p_n$ variables are directly discarded. Then, a standard variable selection procedure is applied to this reduced model, such as penalized least squares (\ref{beta_penalized_linear}) with  $L_1$ penalty (\ref{L_1 penalty}) or SCAD penalty (\ref{SCAD}). In this way, dependence between covariates is reduced before the application of the procedure for variable selection.
	\item In the second stage, a linear model is built when considering the selected variables in the first step and those in their neighbourhood. That is, if $\widehat{S}_1=\{k \in \{1,\dots,w_n\}, \ \widehat{\beta}_k\not=0\}$, the following set of variables is considered in the second step: $\cup_{k\in \widehat{S}_1}\{\mathcal{X}_i(t_{(k-1)q_n+1}), \dots, \mathcal{X}_i(t_{kq_n})\}$. In this way, relevant information which was missed at the first step is taken into account. After that, the same standard variable selection procedure is applied again to this resulting model.
\end{itemize}
The algorithm requires a division of the sample to be carried out in the two stages. The natural choice is to use half of the sample in the first step and the other half in the second step (in some applications that cannot be the optimal option).

The PVS idea has the advantage of being able to extended it to more complex models with a linear component. In \citet{AneirosVieu2015}, the PVS procedure was extended to the bi-functional partial linear model, which is defined as
\begin{equation}
	Y_i=\sum_{j=1}^{p_n}\beta_{j}\mathcal{X}_i(t_j)+m(\zeta_i)+\varepsilon_i, \ i\in\{1,\dots,n\},\label{MFPLM}
\end{equation}
where $\zeta$ denotes a random variable valued on some semimetric space, and $m(\cdot)$ is an unknown smooth functional (the other notation in model (\ref{SLM1}) remains). The idea to apply the PVS procedure to select relevant variables in the linear component of the model (\ref{MFPLM}), is to transform it into a linear model as in (\ref{LMtransformation}). Then, the estimation of coefficients of the resulting linear model can be obtained by the PVS procedure combined with penalized least squares  (\ref{beta_penalized_linear}) with SCAD penalty (\ref{SCAD}).

One of the main advantages of model (\ref{MFPLM}) is that it allows the inclusion of both, pointwise and continuous effects of functional predictors (which was found as profitable in \citet{FerratyHallVieu2010} or in \citet{Kneipetal2016}).  
However, the presence of the nonparametric component could bring interpretability and dimensionality problems in some applications.
For that, in \citet{NovoVieuAneiros2021}, the PVS procedure was extended to a complete semiparametric model, which replaces the nonparametric component of the model (\ref{MFPLM}) by a functional single-index structure $m(\zeta_i)=g(\left<\theta,\zeta_i\right>)$, where $\zeta_i$ belongs to some separable Hilbert space with inner product $\left<\cdot,\cdot\right>$, $\theta$ is an unknown functional parameter and $g(\cdot)$ is a real-valued smooth function to estimate.

The problem of variable selection in this model has two additional difficulties in comparison with models (\ref{SLM1}) and (\ref{MFPLM}): the estimation of the functional parameter $\theta$ is computationally expensive and needs a relatively big sample size. Therefore, in addition to the PVS procedure, \citet{NovoVieuAneiros2021} studied the behaviour of the method that uses only the first step of the PVS procedure and obtained good results (from a theoretical and practical point of view). The reason behind this proposal is to reduce computational cost in situations of very large $p_n$ and improve the behaviour of the PVS procedure in situations of small sample size (with only one stage the division of the sample is not needed). 

Up to now, the PVS procedure was applied to models where the discretized functional objects have linear effect in the response. But it can be applied also to select variables in sparse nonparametric functional modelling. \citet{AneirosVieu2016} studied its application in the model given by the expression
\begin{equation*}
	Y_i=\sum_{j=1}^{p_n}f_j(\mathcal{X}_i(t_j))+\varepsilon_i, \ i\in\{1,\dots,n\},
\end{equation*}
where $f_j(\cdot)$ are unknown smooth real-valued functions and $\varepsilon_i$ is the random error. To estimate models in both stages of the PVS procedure and simultaneously select relevant variables in them, authors use a pilot multivariate additive model procedure for variable selection, such as the one proposed  in \citet{Huangetal2010}, based on approximation of the additive components by truncated
series expansions with B-splines bases, and then apply adaptive group LASSO.

\subsection{Selection of functional covariates}\label{Section2.3}

In models with scalar response, $Y_i, \ i\in\{1,\dots,n\}$, and under linear relationship between response and functional predictors, the selection of functional variables requires dealing with the functional nature of the predictors, say $(\mathcal{X}_{i1}(t),\dots,\mathcal{X}_{iM}(t))$ and its corresponding coefficient functions, say $(\alpha_1(t),\dots,\alpha_M(t))$. In the case of non-relevant variables, the corresponding coefficient function should be estimated as constant $0$ for all $t$ in its domain $\mathcal{I}$.

The application of shrinkage methods for selecting relevant functional covariates involves optimization  in a functional space, so the problem can not be directly addressed. Testing procedures for variable selection also require a reduction of the dimension of functional predictors. For dealing with this inconvenience, some authors follow a group modelling strategy, in advance for sake of brevity, GM strategy, which basically consists in transforming the given model into a grouped linear model (\ref{grouped linear model}). Specifically, this includes the following steps.
\begin{enumerate}
	\item Firstly, it is assumed that functional predictors and its corresponding coefficient functions belong to a separable Hilbert space, so they can be expressed via countable orthonormal basis. Therefore, we can obtain basis expansions of the functional objects. These basis expansions can be truncated in order to contain a finite number of basis functions,  and still offer a good approximation of the functional object. The truncation parameter will be denoted as $u_m$, since it will depend on each functional predictor. To sum up, the functional elements of the model can be expanded in the following way,
	\begin{equation}
		\begin{aligned}
			&\mathcal{X}_{im}(t)=\sum_{r=1}^{\infty}W_{imr}\phi_{mr}(t)\approx\sum_{r=1}^{u_m}W_{imr}\phi_{mr}(t)=\pmb{W}_{im}^{\top}\pmb{\phi}_{m}(t),&\\   &\alpha_m(t)=\sum_{r=1}^{\infty}\beta_{mr}\phi_{mr}(t)\approx\sum_{r=1}^{u_m}\beta_{mr}\phi_{mr}(t)=\pmb{\beta}_m^{\top}\pmb{\phi}_{m}(t), \ i\in\{1,\dots,n\}, \ m\in\{1,\dots,M\}& 
		\end{aligned}
		\label{basis expansions}
	\end{equation}
	where $\pmb{W}_{im}=(W_{im1},\dots,W_{imu_m})^{\top}, \ \pmb{\beta}_m=(\beta_{m1},\dots,\beta_{mu_m})^{\top}$ and $\pmb{\phi}_{m}(t)=(\phi_{m1}(t),\dots,\phi_{mu_m}(t))^{\top}$ (note that both $\pmb{W}_{im}$ and $\pmb{\beta}_m$ are vectors of coefficients, while $\pmb{\phi}_{m}(t)$ are vectors of basis functions). 
	%are vectors of coefficients, and $\pmb{\phi}_{m}(t)=(\phi_{m1}(t),\dots,\phi_{mu_m}(t)))^{\top}$ are vectors of basis functions. 
	For building basis, authors use different approaches, some of them involve known basis functions and others involve empirical basis functions. The first ones include Fourier, B-spline and wavelet bases (see, \citet{RamsaySilverman2005} for a general presentation) or  Gaussian radial bases (see \citet{Andoetal2008}); the last ones include bases based on functional principal components (FPC) (see, for instance, \citet{RamsaySilverman2005} or \citet{HallMullerWang2006}).
	
	\item Therefore, each functional covariate $\mathcal{X}_m(t)$ is in correspondence with a finite set of coefficients $\pmb{\beta}_m$, 
	%=(\beta_{m1},\dots,\beta_{mu_m})^{\top}$ 
	$m\in\{1,\dots,M\}$, that should be treated together in order to select or discard a functional predictor. Then, the model is transformed into a grouped linear model (\ref{grouped linear model}). % the penalized log-likelihood procedure (\ref{beta group}) with group LASSO regularization  (\ref{Group LASSO penalty}), can be applied taking $\ell(\pmb{\beta}_1,\dots,\pmb{\beta}_M)=-\sum_{i=1}^nl_i(g(\pmb{Z}_i^{\top}\pmb{\alpha}+\sum_{m=1}^M\pmb{X}_{mi}^{\top}\pmb{\beta}_m),Y_i)$.
	%Since each group of coefficients $\pmb{\beta}_m$ is in correspondence with a functional covariate, the mechanism allows variable selection of relevant functional variables. 
\end{enumerate}
Combining this strategy in functional linear modelling  with penalized methods,
the optimization problem is reduced to (\ref{beta group}) and group penalties or sparsity-smoothing penalties can be applied.
This combination can be seen in several papers. For instance,  \citet{MatsuiKonishi2011} %trying to discriminate disease from non-disease in the diagnosis of cervical precancer, 
studied the selection of functional covariates in a multiple functional linear model given by the expression %$\mathbb{E}(Y_i|\pmb{Z}_i,\mathcal{X}_{1i},\dots,\mathcal{X}_{Mi})=g^{-1}(\eta_i),$ and
\begin{equation}
	\begin{aligned}
		&Y_i=\sum_{m=1}^M\int_{\mathcal{I}}\alpha_m(t)\mathcal{X}_{im}(t)dt+\varepsilon_i,\ i\in\{1,\dots,n\},&
	\end{aligned}
	\label{MFLM}
\end{equation}
where $\varepsilon_i$ denotes the random error.
%where $Y_i$ is a scalar response, $\mathcal{X}_{mi}(t)$  are functional predictors with domain $\mathcal{I}$ %and $\mathbb{E}(\mathcal{X}_{im}(t))=\mu_m(t)$ 
%for $m\in\{1,\dots,M\}$ and $i\in\{1,\dots,n\}$, %while $\pmb{Z}_i=(Z_{i1},\dots,Z_{ip})$ is a vector of scalar covariates.
%In addition,  %$\pmb{\gamma}=(\gamma_1,\dots,\gamma_p)$ is the vector of coefficients of the scalar covariates, while
%  $\alpha_m(t)$ $(m\in\{1,\dots,M\})$ are coefficient functions. %Finally, $g$ is assumed to be a known injective continuous function.
\citet{MatsuiKonishi2011} used Gaussian bases for constructing basis expansions (\ref{basis expansions}). 
After that, expression (\ref{MFLM}) is transformed:%$\sum_{m=1}^M\int_{\mathcal{I}}\alpha_m(t)\mathcal{X}_{mi}(t)dt=\sum_{m=1}^M\int_{\mathcal{I}}\pmb{W}_{mi}^{\top}\pmb{\phi}_{m}(t)\pmb{\phi}_{m}(t)^{\top}\pmb{\beta}_mdt$. Therefore, using notation $\pmb{\beta}^*=(\pmb{\beta}_1,\dots,\pmb{\beta}_M)^{\top}$, in expression (\ref{beta group}) they take $\ell(\pmb{\beta}^*)=-\sum_{i=1}^nl_i(g(\pmb{Z}_i^{\top}\pmb{\alpha}+\sum_{m=1}^M\pmb{X}_{mi}^{\top}\pmb{\beta}_m),Y_i)$ and penalty (\ref{}) 
\begin{equation}
	Y_i=\sum_{m=1}^M\int_{\mathcal{I}}\pmb{W}_{mi}^{\top}\pmb{\phi}_{m}(t)\pmb{\phi}_{m}(t)^{\top}\pmb{\beta}_mdt+\varepsilon_i=\sum_{m=1}^M\pmb{X}_{im}^{\top}\pmb{\beta}_m+\varepsilon_i, \ i\in\{1,\dots,n\},\label{MFLMB}
\end{equation}
where $\pmb{X}_{im}=(\pmb{W}_{im}^{\top}J_{\phi_m})^{\top}$ and $J_{\phi_m}=\int_{\mathcal{I}} \pmb{\phi}_{m}(t)\pmb{\phi}_{m}(t)^{\top}, \ i\in\{1,\dots,n\}, \ m\in\{1,\dots,M\}$. Therefore, an optimization problem of type (\ref{beta group}) is obtained. So using notation $\pmb{\beta}^*=(\pmb{\beta}_1,\dots,\pmb{\beta}_M)^{\top}$, they take $\ell^*(\pmb{\beta}^*)=\sum_{i=1}^n(Y_i-\sum_{m=1}^M\pmb{X}_{im}^{\top}\pmb{\beta}_m)^2$ and penalty function (\ref{Group SCAD penalty}). Other examples using the GM strategy and penalization methods for selecting relevant functional predictors in model (\ref{MFLM}) are
\citet{Lian2011}  (who studied selection of relevant variables as \citet{MatsuiKonishi2011}, but  %That is, they studied the multiple functional regression model, given by the expression
%\begin{equation}
%Y_i=\sum_{m=1}^M\int_{\mathcal{I}}\alpha_m(t)\mathcal{X}_{mi}(t)dt+\varepsilon_i\ \ (i\in\{1,\dots,n\}),\label{MFLM}
%\end{equation}
%where $Y_i$ is a scalar response and $\varepsilon_i$ is the random error (and the other notations are maintained respect to expression(\ref{SFGLM})).
using  FPC basis expansions in (\ref{basis expansions})) and \citet{Huangetal2016}, who studied robust estimation of model (\ref{MFLM}) using penalized LAD method, combined with FPC expansions in (\ref{basis expansions}) and CAP penalty (\ref{CAP}) with $\gamma_0=1$ and  $\gamma_m=\infty$ for all $m\in\{1,\dots,M\}$.
But the GM strategy was also used to select variables via penalization in more complex models, such as generalized multiple functional linear models
(\citet{ZhuCox2009} studied the selection of functional predictors in a model which also included scalar covariates, using a FPC basis and group LASSO regularization (\ref{Group LASSO penalty}); \citet{Gertheissetal2013} used  B-spline basis expansions and sparsity-smoothness penalties (\ref{sparsity-smoothness penalty})) or multiclass logistic regression for functional data (see \citet{Matsui2014} and \citet{Matsui2014b}).

Since datasets containing both functional and non-functional predictors are very frequent, \citet{Kongetal2016} studied simultaneous variable selection of both functional (following the GM strategy) and scalar covariates. Specifically, they worked with the model given by the expression 
\begin{equation}
	Y_i=\pmb{Z}_i^{\top}\pmb{\gamma}+\sum_{m=1}^M\int_{\mathcal{I}}\alpha_m(t)\mathcal{X}_{im}(t)dt+\varepsilon_i, \  i\in\{1,\dots,n\},\label{PLMFLM}
\end{equation}
where $\pmb{Z}_i=(Z_{i1},\dots,Z_{ip})^{\top}$ is a vector of scalar covariates,
$\pmb{\gamma}=(\gamma_1,\dots,\gamma_p)^{\top}$ is a vector of coefficients and $\varepsilon_i$ denotes the random error. They use FPC to obtain expansions (\ref{basis expansions}). In order to obtain sparse estimators of coefficients in both components of the model, the optimization problem that they solved was a combination of (\ref{beta group}), due to the functional predictors, and (\ref{beta_penalized_linear}) due to the scalar predictors. Specifically, using notation $\pmb{\beta}^*=(\pmb{\beta}_1,\dots,\pmb{\beta}_M)^{\top}$ the optimization problem to solve is
\begin{equation}
	(\hat{\pmb{\beta}}^*,\hat{\pmb{\gamma}})=\arg\min_{\pmb{\beta }^{\ast }\in \mathbb{R}^{u_{1}}\times
		\cdots \times \mathbb{R}^{u_{M}}, \ \pmb{\gamma}\in\mathbb{R}^p} \left(\ell^*(\pmb{\beta}^*,\pmb{\gamma})+nP_{\lambda}^{*}(\pmb{\beta}^*)+nP_{\lambda}(\pmb{\gamma})\right),\label{beta functional scalar}
\end{equation}
where $\ell^*(\pmb{\beta}^*,\pmb{\gamma})=\sum_{i=1}^n(Y_i-\sum_{m=1}^M\pmb{\beta}_{m}^{\top}\pmb{X}_m-\pmb{Z}_ i^{\top}\pmb{\gamma}),$ using notations in (\ref{MFLMB}); \citet{Kongetal2016} use as penalties group SCAD (\ref{Group SCAD penalty}) and SCAD (\ref{SCAD}).
\citet{Maetal2019} extended the  procedure in \citet{Kongetal2016} to quantile regression for functional partially linear regression. In the model they studied, the conditional quantile $\tau$ of the response variable follows expression (\ref{PLMFLM}) and the response variable has linear relation with the conditional quantile. %Specifically, they worked with the model
%\begin{equation}
%Y_i=\mathcal{Q}_{\tau}(Y_i|\mathcal{X}_{1i}(t),\dots,\mathcal{X}_{Mi}(t),\pmb{Z}_i)+\varepsilon_{i\tau} \ \ \ (i\in\{1,\dots,n\}, \ \tau\in(0,1))
%\end{equation}
%where $\mathcal{Q}_{\tau}(\cdot)$ denotes the conditional quantile function for quantile $\tau$, which is given by the expression
%\begin{equation}
%\mathcal{Q}_{\tau}(Y_i|\mathcal{X}_{1i}(t),\dots,\mathcal{X}_{Mi}(t),\pmb{Z}_i)=\pmb{Z}_i^{\top}\pmb{\gamma}+\sum_{m=1}^M\int_{\mathcal{I}}\alpha_m(t)\mathcal{X}_{mi}(t)dt
%\end{equation}
For selecting relevant functional and non-functional variables simultaneously, \citet{Maetal2019} follow the same techniques as \citet{Kongetal2016}.

The idea of using various penalization terms can be extended to other contexts, for instance, when we have two different groups of functional predictors. \citet{Fengetal2021} studied model (\ref{MFLM}) but added a term for taking into account interaction effects between functional predictors. So their purpose was to identify relevant main effects and corresponding
interactions associated with the response variable. For that, they carried out variable selection in both terms of the model (main effects and  interactions) separately. Firstly, they followed the GM strategy, obtaining basis expansions of the functional predictors, coefficient functions corresponding with main effects and those corresponding with interaction effects (they used FPC for the basis). Then, they applied least squares estimation combined with two adaptive group lasso penalties (that is, penalties of form (\ref{Group LASSO penalty}) but added weights in the sum as in (\ref{adaptive lasso penalty})), one for main effects and the other for interactions.

In the previously commented papers, it is assumed that there is linear relationship between the response (or a known function of the response) and the functional predictors. However, this assumption can be restrictive in some contexts and more accurate fits can often be produced by
modelling a nonlinear relationship. For that,
\citet{Fanetal2015} went further using the GM strategy combined with shrinkage methods: they studied  a model with scalar response but nonlinear relationship with the functional predictors. Specifically, they studied  sparse functional additive regression (FAR), given by the relationship
\begin{equation}
	Y_i=\sum_{m=1}^M f_m(\mathcal{X}_{im}(t))+\varepsilon_i, \ i\in\{1,\dots,n\},\label{FAR}
\end{equation}
where the $f_m(\cdot)$ are general nonlinear functionals of $\mathcal{X}_{im}(t), \ m\in\{1,\dots,M\}$. To fit model (\ref{FAR})  and simultaneously select relevant predictors, an optimization problem of type (\ref{beta additive models}) should be solved. Denoting $\pmb{f}^*=(f_1,\dots,f_M)^{\top}$, in this case
$\ell^{**}(\pmb{f}^*)=\sum_{i=1}^n\left(Y_i-\sum_{m=1}^Mf_m(\mathcal{X}_{im}(t))\right)^2$, and instead of using a particular penalization $P^{**}_{\lambda}(\pmb{f}^*)$, \citet{Fanetal2015} explore general penalizations of the form
\begin{equation}P^{**}_{\lambda}(\pmb{f}^*)=\sum_{m=1}^M\rho_{\lambda}\left(\sqrt{\frac{1}{n}\sum_{i=1}^n f_m(\mathcal{X}_{im})^2}\right)\end{equation} (that is, involving $l_2$-norm of the vectors of values of $f_m(\cdot)$ evaluated in the sample), where $\rho_{\lambda}(\cdot)$ is a concave  function. However, given the functional nature of $\mathcal{X}_{im}(t)$, solving problem (\ref{beta additive models}) requires knowing the form of the functionals $f_m(\cdot)$. For that, they specialize the methodology to the linear case, using model (\ref{MFLM}) already widely studied, and to the nonlinear case, studying the multiple functional single-index model taking
\begin{equation}
	f_m(\mathcal{X}_{mi}(t))=g_m\left(\int_{\mathcal{I}}\alpha_m(t)\mathcal{X}_{im}(t)dt\right), \ m\in\{1,\dots,M\},
\end{equation}
where $g_m(t)$ are smooth nonparametric functions.

In this case, the usage of the GM strategy needs the expansions (\ref{basis expansions}) (\citet{Fanetal2015} use orthonormal basis expansions, with dimension independent from the predictor $u_m=u$, but  depending on the sample size $u=u_n$) and also requires obtaining basis expansions for functions $g_m(\cdot)$, that is, $g_m(x)\approx\pmb{h}(x)^{\top}\pmb{\delta}_m$ where $\pmb{h}(x)$ is a basis of dimension $V=V_n$. Then the optimization problem (\ref{beta additive models}) is transformed into (\ref{beta group}) but in this case, the optimization is carried out by minimizing the objective function $\ell^*(\pmb{\beta}^*,\pmb{\delta}^*) + nP_{\lambda}^*(\pmb{\beta}^*,\pmb{\delta}^*)$ with respect to $\pmb{\beta}^*=(\pmb{\beta}_1,\dots,\pmb{\beta}_M)$ and $\pmb{\delta}^*=(\pmb{\delta}_1,\dots,\pmb{\delta}_M)^{\top}$, where $$\ell^*(\pmb{\beta}^*,\pmb{\delta}^*)=\sum_{i=1}^n\left(Y_i-\sum_{m=1}^M\pmb{h}(\pmb{W}_{im}^{\top}\pmb{\beta}_m)^{\top}\pmb{\delta}_m)\right)^2 \text{ and } P_{\lambda}^*(\pmb{\beta}^*,\pmb{\delta}^*)=\sum_{m=1}^M\rho_{\lambda}(\sqrt{(1/n)\sum_{i=1}^n\pmb{h}(\pmb{W}_{im}^{\top}\pmb{\beta}_m)^{\top}\pmb{\delta}_m}).$$

So far we have presented procedures based on penalization techniques for selecting relevant functional variables, but in the literature there exist other options for selecting relevant functional predictors, such as testing procedures. In fact, there is a connection between model testing and variable selection: dropping a variable from the model is equivalent to not reject the null hypothesis that its corresponding parameter is $0$. In this case, the GM strategy can be applied in order to reduce dimensionality of covariates and function-parameters and apply finite-dimensional techniques. \citet{Collazosetal2016} worked with model (\ref{MFLM}) applying significance testing of the functional predictors $\mathcal{X}_{im}(t)$, $m\in\{1,\dots,M\}$. For that, they firstly obtained basis expansions (\ref{basis expansions}), and then formulated the following test, for each $m\in\{1,\dots,M\}$:
\begin{equation}
	H_0: \ \pmb{\beta}_m=\pmb{0}, \ H_1: \ \pmb{\beta}_m\not=\pmb{0}. \label{testBETA}
\end{equation}
The test can be solved via a likelihood ratio-test which compares the residual sum of squares (RSS) obtained under the elimination of the predictor $m$  from the model ($H_0$), with the RSS obtained with the complete model. %Based on $p$-values from testing each predictor, they propose a variable selection method based on false discovery rate or Bonferroni corrections applied on the choice of $p$-values.
Since the test (\ref{testBETA}) is performed for each functional predictor, the $M$ $p$-values obtained should be corrected using Bonferroni correction or the false discovery rate. Then, one selects as influential predictors those covariates, the corrected $p$-value of which leads to the rejection of the null hyphotesis in (\ref{testBETA}).
%The
%final model is selected by using generalized information criteria.\\ %After the application of the algorithms the final model was selected using a generalized information criterion.

Other procedures for variable selection, not based on penalization, were presented in \citet{SmagaMatsui2018}. They worked with model (\ref{MFLM}), and obtained basis expansions (\ref{basis expansions}). After that, they built two algorithms based on the random subspace method proposed in \citet{MielniczukTeisseyre2014}, which explores subsets of variables and measure variable importance using $t$-statistics.

Another way of selecting relevant functional predictors without using a penalization term is the Bayesian approach.  \citet{ZhuCox2010} studied variable selection in a Bayesian functional hierarchical model for classification, to deal
with situations when functional predictors are contaminated
by random batch effects. This model uses expression (\ref{PLMFLM}) with latent response in order  to classify observations from a binary random variable. For the estimation procedure, they assume Gaussian processes for the priors for the parameter-functions and introduce an hyperparameter in them that indicates if the functional variable is selected or not. They used orthonormal basis expansions (\ref{basis expansions}) (in particular, they used FCA basis) to reduce dimensionality of functional objects and transformed the functional posterior sampling problem into a multivariate one, and then applied a hybrid Metropolis–Hastings/Gibbs sampler (see, for instance, \citet{GeorgeMcCulloch1997}) to obtain
posterior samples of the parameters and estimate them (selecting relevant variables at the same time).

A different proposal, not involving penalized methods, can be found in \citet{Febreroetal2019}. They work in the context of general additive regression with scalar response, that is, they work with model (\ref{FAR}) but with the difference that predictors can be of different nature (functional, scalar, multivariate, directional, etc.); in addition, the effect of each predictor, $f_m(\cdot)$, can be linear or nonlinear. For selecting relevant variables in that model, they construct an algorithm based on distance correlation $\mathcal{R}(\cdot,\cdot)$ proposed in \citet{Szekelyetal2007}, which only depends on distances among data. If $X_i$ denotes a predictor of any nature, $\mathcal{R}(X_i,Y_i)=0$ characterizes independence with the response. Therefore, the algorithm starts with a null model and
sequentially selects new variables to be incorporated into the model accordingly with $\mathcal{R}(\cdot,\cdot)$. Specifically, the covariate, that provides the high value for  the distance correlation with the current residual, is chosen and, then, a test of independence based on $\mathcal{R}(\cdot,\cdot)$ is carried out: the variable is candidate to be included into the model only if the null hypotheses of independence is rejected. Each candidate variable is added to the model fixing the effects (linear/nonlinear) of the previously added variables and setting as the effect of the candidate variable the one (linear/nonlinear) that gives rise to the best contribution; then, the model is checked again to see if the addition of this covariate is relevant or not (by means of a generalized likelihood ratio test). The procedure
ends when no more variables can be added to the model because the set of remaining candidates is empty or all the remaining
variables accept the independence null hypothesis of the distance correlation test.\\

\subsection{The functional response case}\label{Section2.4}

It also could be the case that the functional variable is the one that we want to predict, and for that we have a set of scalar covariates, but only few of them are really related with the response. In this framework \citet{Wangetal2007} proposed a model with functional response and time-varying coefficients. Specifically, the model is given by the relationship
\begin{equation}
	Y_i(t)=\sum_{m=1}^M\alpha_m(t)X_{im}+\varepsilon_i(t), \ i\in\{1,\dots,n\},\label{MFLMYfun}
\end{equation}
where $Y_i(t)$ is a functional variable with domain $\mathcal{I}$, $X_{i1},\dots,X_{iM}$ are real covariates, $\alpha_m(t)$ is a function-parameter and $\varepsilon_i(t)$ is a stochastic process corresponding to the random error. For estimating function-parameters, they follow the GM strategy: they expand these coefficient functions using B-spline basis (\ref{basis expansions}) with $u_m=u$, and the same basis functions for all the coefficients; then they use penalized least squares (with the discretized response $Y_i(t_1),\dots, Y_i(t_T)$, that is,  $l^*(\pmb{\beta}^*)=\sum_{i=1}^n\sum_{j=1}^{T}(Y_i(t_j)-\sum_{m=1}^M(\sum_{r=1}^u\beta_{mr}\phi_r(t_j))X_{im})^2$) and propose group SCAD penalty (\ref{Group SCAD penalty}).

\citet{Mingottietal2013} also studied model (\ref{MFLMYfun}). In order to estimate coefficient functions they applied a penalized least squares procedure, obtaining previously B-spline expansions for the coefficient-functions (\ref{basis expansions}) and for the response variable using $u_m=u$ and the same basis for all the parameter functions and for the response, that is, $Y_i(t)\approx\sum_{r=1}^ua_{ri}\phi_r(t)$ and $l^*(\pmb{\beta}^*)=\sum_{i=1}^n\int_{\mathcal{I}}(\sum_{r=1}^ua_{ri}\phi_r(t)-\sum_{m=1}^M(\sum_{r=1}^u\beta_{mr}\phi_r(t))X_{im})^2dt$. In this case, the penalty term proposed, called functional LASSO, is given by the expression
\begin{equation*}
	P_{\lambda}^*(\pmb{\beta}^*)= n\lambda\sum_{m=1}^M\int_{\mathcal{I}}\left\lvert\sum_{r=1}^u\beta_{mr}\phi_r(t)\right\rvert dt.
\end{equation*}
\citet{Mingottietal2013} got advantage of B-spline properties in computations of integrals (see \citet{deBoor2001}).

\citet{HongLian2011} generalize variable selection with LASSO penalty for multiple functional linear model for the case in which both response variable and covariates are functional, but parameters of the model are scalar.

\section{Conclusions and future perspectives}\label{Section3}
As discussed along this review there exists a rich production in variable selection methods for functional regression. Many different techniques have been developed, all of them incorporate in some sense previous ideas from variable selection in mutivariate regression models (most of them, ideas related to LASSO, despite the criticism received by this selector; it is expected that ideas based on other selectors will be adapted in the near future to the functional case). This is an evidence of the needs for bridging gaps between FDA and HDS, and also of the benefits one could get from such crossing of the ideas. The functional production is still very small compared to the dramatically high number of research papers related with this topic in finite-dimensional models, and undoubtly the next few years should lead to many new advances in the domain, and the extension of techniques in the functional setting is expected to continue.% as happened in other fields.  %{\color{blue}{Aqui creo que podemos pensar en incluir una frase (no en la primera version, sino en la version final cuando ya tendremos los informes de referees). Esta frase serviria para anadir los ultimos papeles, sin tener que cambiar nada de la estructura del articulo. Posiblamente los referees van a indicarnos algunos otros papeles. Puede ser tambien el caso que salen otros articulos ... Esta frase seria algo muy corto, asi: Let us point out that since the first writing of this paper we have been aware of a few recent advances (see [], [])}}.\\

As we have commented in the Introduction, the complexity of the data continues to increase day by day. \citet{Muller2016}  talked about ``next generation'' functional data, and provided some speculative notions about the challenges that the area has to overcome in the future (functional data irregularly and sparsely observed, the mix between big data and functional data, the multivariate time domain in biostatistical modelling, etc). In this context, the dimension and the increasing number of predictors become a more serious problem. Undoubtly, there will be the necessity of building new models and, consequently, new  variable selection techniques to treat efficiently this new kind of functional objects.

\section*{Acknowledgments}
The authors are confident on the fact that this volume will contribute (as JMVA did since 50 years) to promote further researches in high (and infinite) dimensional statistics, and they wish to express their sincere gratitude to Professor Dietrich von Rosen for having taken the initiative of this Jubilee Volume and for having invited us in presenting a contribution.

This research was supported by MICINN grant PID2020-113578RB-I00 and by the Xunta de Galicia (Grupos de
Referencia Competitiva ED431C-2020-14 and Centro de Investigación del Sistema Universitario de Galicia
ED431G 2019/01), all of them through the ERDF. The second author also thanks the financial support from the
Xunta de Galicia and the European Union (European Social Fund - ESF), the reference of which is ED481A-
2018/191.

%\section*{References}

% To ensure accuracy, get them from MathSciNet whenever possible. Typeset them with BibTeX using JMVA's style file, \texttt{myjmva.bst}.
%\bibliography{}
\bibliographystyle{plain}
%\begin{thebibliography}
\bibliography{trial}

\begin{thebibliography}{}

\bibitem[Ahmed, 2017]{ahmed2017}
Ahmed, S.~E., editor (2017).
\newblock {\em Big and complex data analysis}, Contributions to Statistics.
  Springer.

\bibitem[Ait-Sa{\"i}di et~al., 2008]{AitSaidietal2008}
Ait-Sa{\"i}di, A., Ferraty, F., Kassa, R., and Vieu, P. (2008).
\newblock Cross-validated estimations in the single-functional index model.
\newblock {\em Statistics}, 42(6):475--494.

\bibitem[Ando et~al., 2008]{Andoetal2008}
Ando, T., Konishi, S., and Imoto, S. (2008).
\newblock Nonlinear regression modeling via regularized radial basis function
  networks.
\newblock {\em Journal of Statistical Planning and Inference},
  138(11):3616--3633.

\bibitem[Aneiros et~al., 2019a]{Aneirosetal2019}
Aneiros, G., Cao, R., Fraiman, R., Genest, C., and Vieu, P. (2019a).
\newblock Recent advances in functional data analysis and high-dimensional
  statistics.
\newblock {\em Journal of Multivariate Analysis}, 170:3--9.

\bibitem[Aneiros et~al., 2019b]{aneiros2019}
Aneiros, G., Cao, R., Fraiman, R., and Vieu, P. (2019b).
\newblock Editorial for the special issue on functional data analysis and
  related topics.
\newblock {\em Journal of Multivariate Analysis}, 170:1--2.

\bibitem[Aneiros et~al., 2015]{AneirosFerratyVieu2015}
Aneiros, G., Ferraty, F., and Vieu, P. (2015).
\newblock Variable selection in partial linear regression with functional
  covariate.
\newblock {\em Statistics}, 49(6):1322--1347.

\bibitem[Aneiros et~al., 2022]{Aneirosetal2022}
Aneiros, G., Horova, I., Huskova, M., and Vieu, P. (2022).
\newblock On functional data analysis and related topics.
\newblock {\em Journal of Multivariate Analysis}, 189:104861.

\bibitem[Aneiros and Vieu, 2014]{AneirosVieu2014}
Aneiros, G. and Vieu, P. (2014).
\newblock Variable selection in infinite-dimensional problems.
\newblock {\em Statistics {\&} Probability Letters}, 94:12--20.

\bibitem[Aneiros and Vieu, 2015]{AneirosVieu2015}
Aneiros, G. and Vieu, P. (2015).
\newblock Partial linear modelling with multi-functional covariates.
\newblock {\em Computational Statistics}, 30(3):647--671.

\bibitem[Aneiros and Vieu, 2016]{AneirosVieu2016}
Aneiros, G. and Vieu, P. (2016).
\newblock Comments on: {P}robability enhanced effective dimension reduction for
  classifying sparse functional data.
\newblock {\em TEST}, 25:27--32.

\bibitem[Berrendero et~al., 2019]{Berrenderoetal2019}
Berrendero, J.~R., Bueno-Larraz, B., and Cuevas, A. (2019).
\newblock An {RKHS} model for variable selection in functional linear
  regression.
\newblock {\em Journal of Multivariate Analysis}, 170:25--45.

\bibitem[Bongiorno et~al., 2014]{bongiorno2014}
Bongiorno, E.~G., Goia, A., Salinelli, E., and Vieu, P. (2014).
\newblock An overview of {IWFOS}'2014.
\newblock In {\em Contributions in Infinite-Dimensional Statistics and Related
  Topics}, pages 1--5. Esculapio, Bologna.

\bibitem[Breiman, 1996]{Breiman1996}
Breiman, L. (1996).
\newblock {Heuristics of instability and stabilization in model selection}.
\newblock {\em The Annals of Statistics}, 24(6):2350 -- 2383.

\bibitem[Cand{\`e}s and Tao, 2007]{CandesTao2007}
Cand{\`e}s, E. and Tao, T. (2007).
\newblock The {D}antzig selector: statistical estimation when p is much larger
  than n.
\newblock {\em Annals of Statistics}, 35:2392--2404.

\bibitem[Collazos et~al., 2016]{Collazosetal2016}
Collazos, J.~A., Dias, R., and Zambom, A.~Z. (2016).
\newblock Consistent variable selection for functional regression models.
\newblock {\em Journal of Multivariate Analysis}, 146:63--71.

\bibitem[Cuevas, 2014]{Cuevas2014}
Cuevas, A. (2014).
\newblock A partial overview of the theory of statistics with functional data.
\newblock {\em Journal of Statistical Planning and Inference}, 147:1--23.

\bibitem[de~Boor, 2001]{deBoor2001}
de~Boor, C. (2001).
\newblock {\em A Practical Guide to Splines}.
\newblock Applied Mathematical Sciences. Springer-Verlag, New York.

\bibitem[Desboulets, 2018]{Desboulets2018}
Desboulets, L. D.~D. (2018).
\newblock A review on variable selection in regression analysis.
\newblock {\em Econometrics}, 6(4).

\bibitem[Efron et~al., 2004]{Efronetal2004}
Efron, B., Hastie, T., Johnstone, I., and Tibshirani, R. (2004).
\newblock Least angle regression.
\newblock {\em Annals of Statistics}, 32:407--499.

\bibitem[Efroymson, 1960]{Efroymson1960}
Efroymson, M.~A. (1960).
\newblock Multiple regression analysis.
\newblock In Ralston, A. and Wilf, H.~S., editors, {\em Mathematical Methods
  for Digital Computers}, New York. Wiley.

\bibitem[Fan and Gijbels, 1996]{FanGijbels1996}
Fan, J. and Gijbels, I. (1996).
\newblock {\em Local Polynomial Modelling and its Applications}.
\newblock Monographs on Statistics and Applied Probability 66. Routledge.

\bibitem[Fan and Li, 2001]{FanLi2001}
Fan, J. and Li, R. (2001).
\newblock Variable selection via nonconcave penalized likelihood and its oracle
  properties.
\newblock {\em Journal of the American Statistical Association},
  96(456):1348--1360.

\bibitem[Fan and Lv, 2008]{FanLv2008}
Fan, J. and Lv, J. (2008).
\newblock Sure independence screening for ultrahigh dimensional feature space.
\newblock {\em Journal of the Royal Statistical Society: Series B (Statistical
  Methodology)}, 70(5):849--911.

\bibitem[Fan and Lv, 2010]{FanLv2010}
Fan, J. and Lv, J. (2010).
\newblock A selective overview of variable selection in high dimensional
  feature space.
\newblock {\em Statistica Sinica}, 20(1):101--148.

\bibitem[Fan and Peng, 2004]{FanPeng2004}
Fan, J. and Peng, H. (2004).
\newblock Nonconcave penalized likelihood with a diverging number of
  parameters.
\newblock {\em Annals of Statistics}, 32(3):928--961.

\bibitem[Fan et~al., 2015]{Fanetal2015}
Fan, Y., James, G.~M., and Radchenko, P. (2015).
\newblock {Functional additive regression}.
\newblock {\em The Annals of Statistics}, 43(5):2296--2325.

\bibitem[Febrero-Bande et~al., 2019]{Febreroetal2019}
Febrero-Bande, M., Gonz{\'a}lez-Manteiga, W., and de~la Fuente, M.~O. (2019).
\newblock Variable selection in functional additive regression models.
\newblock {\em Computational Statistics}, 34:469--487.

\bibitem[Feng et~al., 2021]{Fengetal2021}
Feng, S., Zhang, M., and Tong, T. (2021).
\newblock Variable selection for functional linear models with strong heredity
  constraint.
\newblock {\em Annals of the Institute of Statistical Mathematics}.

\bibitem[Ferraty et~al., 2010]{FerratyHallVieu2010}
Ferraty, F., Hall, P., and Vieu, P. (2010).
\newblock Most-predictive design points for functional data predictors.
\newblock {\em Biometrika}, 97(4):807--824.

\bibitem[Ferraty and Vieu, 2006]{FerratyVieu2006}
Ferraty, F. and Vieu, P. (2006).
\newblock {\em Nonparametric Functional Data Analysis, Theory and Practice}.
\newblock Springer Series in Statistics. Springer-Verlag, New York.

\bibitem[Frank and Friedman, 1993]{FrankFriedman1993}
Frank, I.~E. and Friedman, J.~H. (1993).
\newblock A statistical view of some chemometrics regression tools.
\newblock {\em Technometrics}, 35(2):109--135.

\bibitem[Furnival and Wilson, 1974]{FurnivalWilson1974}
Furnival, G.~M. and Wilson, R.~W. (1974).
\newblock Regressions by leaps and bounds.
\newblock {\em Technometrics}, 16(4):499--511.

\bibitem[George and McCulloch, 1997]{GeorgeMcCulloch1997}
George, E.~I. and McCulloch, R.~E. (1997).
\newblock Approaches for {B}ayesian variable selection.
\newblock {\em Statistica Sinica}, 7(2):339--373.

\bibitem[Gertheiss et~al., 2013]{Gertheissetal2013}
Gertheiss, J., Maity, A., and Staicu, A.~M. (2013).
\newblock Variable selection in generalized functional linear models.
\newblock {\em Stat}, 2(1):86--101.

\bibitem[Goia and Vieu, 2016]{GoiaVieu2016}
Goia, A. and Vieu, P. (2016).
\newblock An introduction to recent advances in high/infinite dimensional
  statistics.
\newblock {\em Journal of Multivariate Analysis}, 146:1--6.

\bibitem[Hall et~al., 2006]{HallMullerWang2006}
Hall, P., Müller, H.-G., and Wang, J.-L. (2006).
\newblock {Properties of principal component methods for functional and
  longitudinal data analysis}.
\newblock {\em The Annals of Statistics}, 34(3):1493--1517.

\bibitem[Hastie et~al., 2009]{Hastieetal2009}
Hastie, T., Tibshirani, R., and Friedman, J. (2009).
\newblock Linear methods for regression.
\newblock In {\em The Elements of Statistical Learning}, pages 43--99. Springer
  Series in Statistics, New York.

\bibitem[Hong and Lian, 2011]{HongLian2011}
Hong, Z. and Lian, H. (2011).
\newblock Inference of genetic networks from time course expression data using
  functional regression with lasso penalty.
\newblock {\em Communications in Statistics - Theory and Methods},
  40(10):1768--1779.

\bibitem[Huang et~al., 2008]{Huangetal2008}
Huang, J., Horowitz, J.~L., and Ma, S. (2008).
\newblock Asymptotic properties of bridge estimators in sparse high-dimensional
  regression models.
\newblock {\em The Annals of Statistics}, 36(2):587--613.

\bibitem[Huang et~al., 2010]{Huangetal2010}
Huang, J., Horowitz, J.~L., and Wei, F. (2010).
\newblock {Variable selection in nonparametric additive models}.
\newblock {\em The Annals of Statistics}, 38(4):2282--2313.

\bibitem[Huang et~al., 2016]{Huangetal2016}
Huang, L., Zhao, J., Wang, H., and Wang, S. (2016).
\newblock Robust shrinkage estimation and selection for functional multiple
  linear model through lad loss.
\newblock {\em Computational Statistics {\&} Data Analysis}, 103:384--400.

\bibitem[Ke et~al., 2014]{Keetal2014}
Ke, Z.~T., Jin, J., and Fan, J. (2014).
\newblock Covariate assisted screening and estimation.
\newblock {\em The Annals of Statistics}, 42(6):2202--2242.

\bibitem[Kneip et~al., 2016]{Kneipetal2016}
Kneip, A., Poß, D., and Sarda, P. (2016).
\newblock {Functional linear regression with points of impact}.
\newblock {\em The Annals of Statistics}, 44(1):1--30.

\bibitem[Kneip and Sarda, 2011]{KneipSarda2011}
Kneip, A. and Sarda, P. (2011).
\newblock Factor models and variable selection in high-dimensional regression
  analysis.
\newblock {\em The Annals of Statistics}, 39(5).

\bibitem[Kong et~al., 2016]{Kongetal2016}
Kong, D., Xue, K., Yao, F., and Zhang, H.~H. (2016).
\newblock {Partially functional linear regression in high dimensions}.
\newblock {\em Biometrika}, 103(1):147--159.

\bibitem[Leng et~al., 2006]{Lengetal2006}
Leng, C., Lin, Y., and Wahba, G. (2006).
\newblock A note on the {Lasso} and related procedures in model selection.
\newblock {\em Statistica Sinica}, 16(4):1273--1284.

\bibitem[Li et~al., 2012]{Lietal2012}
Li, R., Zhong, W., and Zhu, L. (2012).
\newblock Feature screening via distance correlation learning.
\newblock {\em Journal of the American Statistical Association},
  107(499):1129--1139.

\bibitem[Lian, 2011]{Lian2011}
Lian, H. (2011).
\newblock Shrinkage estimation and selection for multiple functional
  regression.
\newblock {\em Statistica Sinica}, (23):51--74.

\bibitem[Ma et~al., 2019]{Maetal2019}
Ma, H., Li, T., Zhu, H., and Zhu, Z. (2019).
\newblock Quantile regression for functional partially linear model in
  ultra-high dimensions.
\newblock {\em Computational Statistics {\&} Data Analysis}, 129:135--147.

\bibitem[Matsui, 2014]{Matsui2014}
Matsui, H. (2014).
\newblock Variable and boundary selection for functional data via multiclass
  logistic regression modeling.
\newblock {\em Computational Statistics {\&} Data Analysis}, 78:176--185.

\bibitem[Matsui, 2019]{Matsui2014b}
Matsui, H. (2019).
\newblock Sparse group lasso for multiclass functional logistic regression
  models.
\newblock {\em Communications in Statistics - Simulation and Computation},
  48(6):1784--1797.

\bibitem[Matsui and Konishi, 2011]{MatsuiKonishi2011}
Matsui, H. and Konishi, S. (2011).
\newblock Variable selection for functional regression models via the {$L_1$}
  regularization.
\newblock {\em Computational Statistics {\&} Data Analysis}, 55(12):3304--3310.

\bibitem[McKeague and Sen, 2010]{McKeagueSen2010}
McKeague, I.~W. and Sen, B. (2010).
\newblock {Fractals with point impact in functional linear regression}.
\newblock {\em The Annals of Statistics}, 38(4):2559--2586.

\bibitem[Meier et~al., 2009]{Meieretal2009}
Meier, L., van~de Geer, S., and Bühlmann, P. (2009).
\newblock {High-dimensional additive modeling}.
\newblock {\em The Annals of Statistics}, 37(6B):3779--3821.

\bibitem[Meinshausen and B{\"u}hlmann, 2006]{MeinshausenBuhlmann2006}
Meinshausen, N. and B{\"u}hlmann, P. (2006).
\newblock {High-dimensional graphs and variable selection with the Lasso}.
\newblock {\em The Annals of Statistics}, 34(3):1436 -- 1462.

\bibitem[Mielniczuk and Teisseyre, 2014]{MielniczukTeisseyre2014}
Mielniczuk, J. and Teisseyre, P. (2014).
\newblock Using random subspace method for prediction and variable importance
  assessment in linear regression.
\newblock {\em Computational Statistics {\&} Data Analysis}, 71:725--742.

\bibitem[Mingotti et~al., 2013]{Mingottietal2013}
Mingotti, N., Lillo~Rodríguez, R.~E., and Romo~Urroz, J. (2013).
\newblock {Lasso variable selection in functional regression}.
\newblock Statistics and Econometrics Series 13 Working paper 13--14,
  Universidad Carlos III de Madrid.

\bibitem[Müller, 2016]{Muller2016}
Müller, H.-G. (2016).
\newblock Peter {H}all, functional data analysis and random objects.
\newblock {\em The Annals of Statistics}, 44(5):1867--1887.

\bibitem[Novo et~al., 2021a]{NovoAneirosVieu2020}
Novo, S., Aneiros, G., and Vieu, P. (2021a).
\newblock Sparse semiparametric regression when predictors are mixture of
  functional and high-dimensional variables.
\newblock {\em TEST}, 30:481--504.

\bibitem[Novo et~al., 2021b]{NovoVieuAneiros2021}
Novo, S., Vieu, P., and Aneiros, G. (2021b).
\newblock Fast and efficient algorithms for sparse semiparametric bi-functional
  regression.
\newblock {\em Australian and New Zealand Journal of Statistics}, 63:606--638.

\bibitem[O'Hara and Sillanp{\"a}{\"a}, 2009]{Oharaetal2009}
O'Hara, R.~B. and Sillanp{\"a}{\"a}, M.~J. (2009).
\newblock {A review of Bayesian variable selection methods: what, how and
  which}.
\newblock {\em Bayesian Analysis}, 4(1):85--117.

\bibitem[Ramsay and Silverman, 2005]{RamsaySilverman2005}
Ramsay, J.~O. and Silverman, B. (2005).
\newblock {\em Functional Data Analysis}.
\newblock Springer Series in Statistics. Springer-Verlag, New York, 2nd
  edition.

\bibitem[Sangalli, 2018]{Sangalli2018}
Sangalli, L.~M. (2018).
\newblock The role of statistics in the era of big data.
\newblock {\em Statistics {\&} Probability Letters}, 136:1--3.

\bibitem[Smaga and Matsui, 2018]{SmagaMatsui2018}
Smaga, L. and Matsui, H. (2018).
\newblock A note on variable selection in functional regression via random
  subspace method.
\newblock {\em Statistical Methods {\&} Applications}, 27:455--477.

\bibitem[Székely et~al., 2007]{Szekelyetal2007}
Székely, G.~J., Rizzo, M.~L., and Bakirov, N.~K. (2007).
\newblock {Measuring and testing dependence by correlation of distances}.
\newblock {\em The Annals of Statistics}, 35(6):2769--2794.

\bibitem[Tibshirani, 1996]{Tibshirani1996}
Tibshirani, R. (1996).
\newblock Regression shrinkage and selection via the {Lasso}.
\newblock {\em Journal of the Royal Statistical Society: Series B},
  58:267--288.

\bibitem[Vieu, 2018]{Vieu2018}
Vieu, P. (2018).
\newblock On dimension reduction models for functional data.
\newblock {\em Statistics {$\&$} Probability Letters}, 136:134--138.

\bibitem[Wang et~al., 2007a]{Wang2etal2007}
Wang, H., Li, G., and Jiang, G. (2007a).
\newblock Robust regression shrinkage and consistent variable selection through
  the lad-lasso.
\newblock {\em Journal of Business {\&} Economic Statistics}, 25(3):347--355.

\bibitem[Wang et~al., 2007b]{Wangetal2007}
Wang, L., Chen, G., and Li, H. (2007b).
\newblock {Group SCAD regression analysis for microarray time course gene
  expression data}.
\newblock {\em Bioinformatics}, 23(12):1486--1494.

\bibitem[Weisberg, 1980]{Weisberg1980}
Weisberg, S. (1980).
\newblock {\em Applied Linear Regression}.
\newblock Wiley, New York.

\bibitem[Yuan and Lin, 2006]{YuanLin2006}
Yuan, M. and Lin, Y. (2006).
\newblock Model selection and estimation in regression with grouped variables.
\newblock {\em Journal of the Royal Statistical Society: Series B (Statistical
  Methodology)}, 68(1):49--67.

\bibitem[Zhao et~al., 2009]{Zhaoetal2009}
Zhao, P., Rocha, G., and Yu, B. (2009).
\newblock {The composite absolute penalties family for grouped and hierarchical
  variable selection}.
\newblock {\em The Annals of Statistics}, 37(6A):3468--3497.

\bibitem[Zhu and Cox, 2009]{ZhuCox2009}
Zhu, H. and Cox, D.~D. (2009).
\newblock A functional generalized linear model with curve selection in
  cervical pre-cancer diagnosis using fluorescence spectroscopy.
\newblock In Rojo, J., editor, {\em Optimality: The Third Erich L. Lehmann
  Symposium}, volume~57, pages 173--189.

\bibitem[Zhu et~al., 2010]{ZhuCox2010}
Zhu, H., Vannucci, M., and Cox, D.~D. (2010).
\newblock A {B}ayesian hierarchical model for classification with selection of
  functional predictors.
\newblock {\em Biometrics}, 66(2):463--473.

\bibitem[Zou, 2006]{Zou2006}
Zou, H. (2006).
\newblock The adaptive {Lasso} and its oracle properties.
\newblock {\em Journal of the American Statistical Association},
  101(476):1418--1429.

\bibitem[Zou and Hastie, 2005]{ZouHastie2005}
Zou, H. and Hastie, T. (2005).
\newblock Regularization and variable selection via the elastic net.
\newblock {\em Journal of the Royal Statistical Society: Series B (Statistical
  Methodology)}, 67(2):301--320.

\end{thebibliography}
%\end{document}

\end{document}